\documentstyle[12pt,aaspp4]{article} 
\def\tempest%
{\begin{array}{ccc} 
1 & 1 & 1 \\ 
1 & 1 & 1 \\ 
4 & 3 & 8 
\end{array}} 
\def\E{{\rm E}}
\def\btheta{\hbox{$\theta\hskip-6.3pt\theta$}}
\def\balpha{\hbox{$\alpha\hskip-7.0pt\alpha$}}
\def\bmu{\hbox{$\mu\hskip-7.5pt\mu$}}
\def\kms{{\rm km}\,{\rm s}^{-1}}

\begin{document}

\title{Photometric Microlens Parallaxes with 
SIM}
\author 
{Andrew Gould
and
Samir Salim}
\affil{Ohio State University, Department of Astronomy, Columbus, OH 43210} 
\affil{E-mail: gould@astronomy.ohio-state.edu, samir@astronomy.ohio-state.edu} 
\begin{abstract} 

	Astrometric measurements of microlensing events can in principle 
determine both the ``parallax'' $\tilde r_\E$ and the ``proper motion'' $\mu$ 
of an individual event which (combined with the Einstein time scale $t_\E$) 
in turn yield the mass, distance, and transverse velocity of the lens.  We 
show, however, that the parallax measurements are generically several orders
of magnitude less precise than the proper-motion measurements.  Fortunately,
astrometric measurements by the Space Interferometry Mission (SIM)
are simultaneously {\it photometric} measurements, and since SIM
will be in solar orbit, these allow SIM to be used as a classical 
(photometric) parallax satellite.  We show that SIM photometric 
parallaxes are of comparable precision to its 
astrometric proper-motion measurements. For 
$I=15$ bulge stars, complete
solutions with $\sim 5\%$ accuracy in mass, distance, and transverse velocity
can be obtained from about 5 hours of 
observation, 100 to 10,000 times shorter than would be required for a
purely astrometric solution of similar precision.  
Thus, it should be
possible to directly measure the mass functions of both the bulge
and the inner disk (including both dark and luminous objects)  with only a few 
hundred hours of SIM observations.

\keywords{astrometry -- Galaxy: stellar content -- gravitational lensing
-- Magellanic Clouds} 
\end{abstract} 
\newpage

\section{Introduction} 

	Boden, Shao \& Van Buren (1998) have shown that it is in principle 
possible to obtain complete solutions for microlensing events from a
series of astrometric measurements using the {\it Space Interferometry
Mission} or possibly ground-based interferometers.   This would be
extremely important if practical because two major questions that
are difficult to answer on the basis of present-day data could then
be easily resolved.

	First, after almost a decade of observations, the nature of the 
events currently being detected toward
the Large Magellanic Cloud (LMC) by the MACHO (Alcock et al.\ 1997b)
and EROS (Aubourg et al.\ 1993) collaborations are a complete mystery.  
On the one
hand, the observed optical depth $\tau\sim 2\times 10^{-7}$ is an order
of magnitude higher than expected from known populations of stars.  On the
other hand, if the lenses lie in the Galactic halo and so comprise of
order half the dark matter, then their masses (inferred from the event
time scales and kinematic models of the halo) are of order half a solar
mass.  Thus, the objects could not be made of hydrogen or they would have
easily been discovered from star counts (Alcock et al.\ 1997b and references
therein).  Direct measurements of the mass and distance of the lenses
would unambigously resolve this question.

	Second, there appears to be a large excess of short time scale events
toward the Galactic bulge relative to what would be expected if bulge stars
had a mass function similar to that seen in the solar neighborhood
(Han \& Gould 1996).  The events would easily be explained if the mass function
were more steeply rising toward low masses (Zhao, Spergel, \& Rich 1995;
Han 1997), but recent observations of the bulge by
Holtzman et al.\ (1998) show that the bulge luminosity function is very
similar to the local one.  By now hundreds of events have been discovered
toward the bulge although only about 50 have been published (Udalski et al.\
1994; Alcock et al.\ 1997a).  If individual masses, positions, and velocities
of even 10\% of these could be measured, our knowledge of the bulge 
population (both dark and luminous) would be dramatically increased.

	In addition, the PLANET (Albrow et al.\ 1999) and MPS (Rhie et al.\
1999) collaborations are currently searching for planetary systems by
closely monitoring ongoing microlensing events seen toward the Galactic bulge.
Ordinarily, these observations can yield only the planet/star mass ratio
and their projected separation in units of the Einstein radius of the lens
(Mao \& Paczy\'nski 1991; Gould \& Loeb 1992).
Complete solutions of the event would enable one to translate these quantities
into planet masses and physical projected separations.

	At present, the only quantity routinely measured for all events
is the Einstein time scale, $t_\E$, which is a complicated combination of the
physical parameters that one would like to know,
\begin{equation}
t_\E = {\theta_E\over \mu},  \qquad \mu = {v\over D_{\rm ol}}
\label{eqn:einstime}
\end{equation}
where $v$ is the transverse speed of the lens relative to the observer-source
line of sight, $\mu$ is the proper motion, and $\theta_E$ is the angular
Einstein radius,
\begin{equation}
\theta_{\rm E}= \biggl({4G M\over D c^2}\biggr)^{1/2},\qquad D\equiv 
{D_{\rm ol}D_{\rm os}\over D_{\rm ls}}.
\label{eqn:thetae}
\end{equation}
Here, $M$ is the mass of the lens, and 
$D_{\rm ol}$, $D_{\rm os}$, and  $D_{\rm ls}$ are the distances between the
observer, lens, and source.  There are numerous ideas on how to get additional
information about individual events, but these often require special
circumstances.  For example, if the source crosses a caustic in the lens
geometry, then it is possible to measure the proper motion $\mu$, and so
from equation (\ref{eqn:einstime}), the Einstein radius (Gould 1994a;
Nemiroff \& Wickramasinghe 1994; Witt \& Mao 1994).  In fact, a variant
of this technique has recently been used to measure the proper motion of
a lens seen toward the Small Magellanic Cloud (SMC) and so demonstrate that
the lens almost certainly resides in the SMC (Afonso et al.\ 1998;
Albrow et al.\ 1999; Alcock et al.\ 1999; Udlaski et al.\ 1998; Rhie et al.\
1999).  However, such caustic crossing events are rare and the great
majority of them are binaries (and hence may not be representative of the
lens population as a whole).

	A second type of information can come from parallax measurements.
If the event is sufficiently long, then the normal light curve is distorted
by the accelerated motion of the Earth about the Sun allowing one to
measure the Einstein radius projected onto the observer plane, $\tilde r_\E$,
\begin{equation}
\tilde r_E = D\theta_\E,\label{eqn:retilde}
\end{equation}
Gould (1992).  Several parallaxes have been measured for bulge events 
(Alcock et al.\ 1995; private communication D.\ Bennett 1998), but all for
events that are substantially longer than typical.  It would be possible
to routinely measure the parallaxes of microlensing events by launching a
satellite into solar orbit (Refsdal 1966; Gould 1995b).  The event would
have a different time of maximum magnification, $t_0$, and different
impact parameter, $\beta$,
as seen from the Earth and the satellite.  From the differences in
these quantities, $\Delta t_0$ and $\Delta \beta$, and using the known
Earth-satellite separation, $d_{\oplus-s}$, and known angle $\gamma$ between
the line of sight and the Earth-satellite vector, one
could reconstruct both the size of the projected Einstein ring, $\tilde r_\E$,
and the direction of
motion, $\phi$, relative to the satellite-Earth vector,
\begin{equation}
\tilde r_\E = {d_{\oplus-s}|\sin\gamma|
\over[(\Delta\beta)^2+(\Delta t_0/t_\E)^2]^{1/2}},
\qquad \tan\phi = {\Delta\beta\over\Delta t_0}
\label{eqn:parallaxeq}
\end{equation}

	A rather technical but in the present context very important
point is that it is significantly easier to measure the difference in times of
maximum, $\Delta t_0$, than it is to measure the difference in impact
parameters, $\Delta \beta$.  There are two interrelated reasons for
this which are investigated in detail by Gould (1994b; 1995b), 
Boutreux \& Gould
(1996), and Gaudi \& Gould (1997).  First, the sign of the impact parameter,
$\beta$,
measured by a single observer is intrinsically ambiguous because the light
curve contains no information about the side of the lens on which the source 
passes.  Hence, from the two individual impact-parameter measurements,
$\beta_\oplus$ and $\beta_s$, it is possible to reconstruct four different
values of $\Delta \beta=\pm(\beta_\oplus \pm \beta_s)$.  Second, one must
determine $\beta$ and $t_0$ from the light curve simultaneously with three
other parameters, $t_E$, $F_0$, and $F_b$, the latter two being the
fluxes from the source star and any unlensed background light that is
blended into the photometric aperture of the source.  While $t_0$ is
virtually uncorrelated with any of these three other parameters, $\beta$
is highly correlated with all of them, in particular with $F_b$.  Hence
$\beta$ (and so $\Delta\beta$) is more poorly measured than $t_0$
(and $\Delta t_0$).  While it is
possible to break the four-fold discrete degeneracy, this requires measurement
of a higher-order effect.

	No dedicated parallax satellite is currently planned.  However,
the Space Infrared Telescope Facility (SIRTF) could be used to measure
parallaxes of at least some events.  Because SIRTF makes its
measurements in passbands that are inaccessible from the ground, the
relative blending between the Earth and satellite is completely unconstrained,
so measurement of $\Delta \beta$ is not simply difficult, it is 
virtually impossible.
Nevertheless, if $\Delta t_0$ is well constrained by Earth-satellite
observations, then it is possible to determine $\Delta \beta$ from
vigorous ground-based observations (Gould 1999a).

	In the best of all possible worlds, one would measure both 
$\theta_\E$ and $\tilde r_\E$.  These (together with the routinely
measured $t_\E$ and the approximately known source distance) would
then yield a complete solution for $M$, $D_{\rm ol}$, and $v$.
(e.g. Gould 1995c).  For example,
\begin{equation}
M = {c^2\over 4 G}\theta_E\tilde r_\E.\label{eqn:msolve}
\end{equation}
At present, this is possible by ground-based measurements only for certain 
rare classes of events (Hardy \& Walker 1995; Gould \& Andronov 1999; Gould 
1997).  If there were a parallax satellite, then it would also be possible
for those rare events which happened to be accessible to proper-motion
measurement.  However, astrometric microlensing opens the possibility,
at least in principle, that such complete
measurements might be made for a large unbiased sample of events in the
future.

\section{Astrometric Microlensing: Promise and Limitations}

	As Boden et al.\ (1998) discuss, astrometric
measurements are sensitive to two distinct effects.  First, the
center of lensed light from the source is displaced from the actual
position of the source by, 
\begin{equation}
\delta\btheta = {{\bf u}_\odot\over u^2_\odot + 2}\theta_{\rm E}
\label{eqn:deltathetamu}
\end{equation}
where ${\bf u}_\odot\equiv \bmu(t-t_0)/\theta_\E$ 
is the projected position of the source relative to the lens
in units of angular Einstein radius assuming rectilinear motion as would
be observed from the Sun.  That is,
\begin{equation}
u_\odot(t) = \biggl[{(t-t_0)^2\over t_\E^2}+\beta^2\biggr]^{1/2}
\label{eqn:uoft}
\end{equation}
This deviation traces out an 
ellipse with semi-major and semi-minor axes,
\begin{equation}
\theta_a = {1\over 2(\beta^2 + 2)^{1/2}}\theta_\E,\qquad
\theta_b = {\beta\over 2(\beta^2 + 2)}\theta_\E.
\label{eqn:thetaab}
\end{equation}
The major axis is aligned with the direction of motion of the lens
relative to the source.  Hence, by measuring this effect, one can solve
for both $\theta_\E$ and the direction of motion.  A measurement
of $\theta_\E$ is often called a ``proper motion'' measurement because,
from equation (\ref{eqn:einstime}) it can be combined with the known
Einstein time scale to yield the magnitude of the proper motion.  However,
in the case of astrometric measurements, it also yields the direction, $\phi$,
and so the full vector proper motion, $\bmu$.

	The second effect is a parallax deviation caused by motion of the
Earth about Sun.  The exact formula for the combined parallax and proper
motion effect can be found by substituting 
\begin{equation}
{\bf u}_\oplus = {\bf u}_\odot -
{{\bf \hat n}\times {\bf \hat n}\times {\bf a}\over \tilde r_\E}
\label{eqn:pardev}
\end{equation}
into equation (\ref{eqn:deltathetamu}).  Here $\bf \hat n$ is the unit vector
in the direction of the source, and ${\bf a}$ is the position vector 
of the Earth relative to the Sun.  Thus, the magnitude of the perturbative
term is $\sim {\rm AU}/\tilde r_\E$, which might be $\sim 10\%$--30\% for 
typical lensing events.  This would seem to imply that one could determine the
parallax $(\tilde r_\E)$ about 10\% to 30\% as accurately as the proper
motion ($\theta_\E$) for the same set of measurements.  

	Unfortunately, the situation is not so favorable.  The perturbation
in equation (\ref{eqn:pardev}) is not directly observable because there
are no comparison observations from the Sun.  Consider the limit
$t_\E\ll {\rm yr}/2\pi$ which is typical for events seen toward the Galactic
bulge.  In this case, the Earth's velocity would barely change during the
event or even for the first few $t_\E$ after it.  One would then see the
same ellipse as described by equation (\ref{eqn:thetaab}), but with a different
$\beta$ from the one that would have been seen from the Sun.  Ninety days
after the event, the direction of apparent source motion would have changed
by an angle $v_\oplus/\tilde v$ where $\tilde v\equiv \tilde r_\E/t_\E$ is
the speed of the lens projected onto the observer plane, and 
$v_\oplus=30\,\kms$ is the speed of the Earth.  According to
equation 
(\ref{eqn:deltathetamu}), this would introduce an astrometric deviation
of order $(v_\oplus/\tilde v)(t_\E/90\,{\rm day})\theta_\E$.  Since $\theta_E$,
is essentially determined from the major axis of the ellipse (eq.\
\ref{eqn:thetaab}), which is approximately $2^{-1/2}\theta_\E$,
the relative size of the astrometric parallax and proper-motion measurement
errors is roughly given by 
\begin{equation}
\Gamma_{\rm ast/ast}\equiv {\sigma_{\tilde r_\E,\rm ast}/\tilde r_\E\over
\sigma_{\theta_\E,\rm ast}/\theta_\E}
\sim{\tilde v\over v_\oplus}\,{60\,{\rm day}\over t_\E}.
\label{eqn:relerrsast}
\end{equation}
In fact, equation (\ref{eqn:relerrsast}) is too optimistic in that it
implicitly assumes that the direction of the major axis of the ellipse
can be determined with infinite precision.  As we will show in \S\ 5,
this is very far from the case.   Hence, the true ratio of errors is generally
larger than implied by equation (\ref{eqn:relerrsast}).
 For typical bulge-bulge lensing events,  $\tilde v\sim 800\,\kms$.
 For typical halo events seen toward the LMC, $\tilde v\sim 300\,\kms$.
 For lenses in the LMC, $\tilde v$ is a factor 3 to 10 higher yet.  

	The above analysis implies that astrometric microlens parallax 
measurements are several orders of magnitude less 
accurate than proper-motion 
measurements.  This would not present much of a problem if very accurate 
proper motion measurements could be made with a modest amount of observing
time.  However, as we will show in \S\ 5, even for bright ($I\sim 15$)
sources seen toward the bulge, proper-motion measurements accurate to 5\%
require about 5 hours of observations.  Very few events seen toward the LMC
are brighter than $V\sim 20$ and therefore $\sim 40$ times more observing
time is needed to achieve the same precision.  Hence, this analysis appears 
to imply that no more than a few accurate microlens parallaxes could be
obtained in any reasonable observing program.

\section{Photometric microlensing parallaxes with SIM}

	SIM is not designed to do photometry, and it would seem
completely hairbrained to waste this precision astrometric instrument on
measurements that could be done more efficiently from the ground using
telescopes with collecting areas that are several orders of magnitude larger.
Nevertheless, two unrelated factors combine to make SIM the
ideal device to measure microlens parallaxes photometrically (rather
than astrometrically).

\subsection{Photometry with SIM}

	First, SIM works by {\it counting photons} as a function
of position in the interference pattern in order to find the centroid of
the central fringe.  The photons are distributed in this fringe as
$NF(\theta)d\theta$, where $N$ is the total number of photons in the
central fringe, 
\begin{equation}
F(\theta) = {1\over \pi\theta_f}
\cos^2\biggl({\theta\over 2\theta_f}\biggr),\qquad
\theta_f \equiv {\lambda\over 2\pi d},
\label{eqn:foftheta}
\end{equation}
$d$ is the distance between the mirrors, and $\lambda$ is the
wavelength of the light.  The astrometric precision is given by,
(e.g.\ Gould 1995a),
\begin{equation}
\sigma_\theta = N^{-1/2}
\biggl[\int d \theta F(\theta)
\biggl({d\ln F\over d\theta}\biggr)^2\biggr]^{-1/2}
= N^{-1/2}\theta_f,
\label{eqn:sigtheta}
\end{equation}
and hence the fractional photometric precision ($\sigma_{\rm ph}=N^{-1/2}$)
is related to the astrometric precision by
\begin{equation}
\sigma_{\rm ph} = {\sigma_\theta\over \theta_f}.
\label{eqn:sigrel}
\end{equation}

\subsection{One-dimensional photometric parallaxes}

	Second, as discussed in the introduction, the real problem in
obtaining microlens parallaxes photometrically is that the microlens
parallaxes are inherently two-dimensional.  In effect, by measuring
$\Delta t_0$, one determines $\cos\phi/\tilde r_\E$, and by measuring
$\Delta \beta$, one determines $\sin\phi/\tilde r_\E$, where $\phi$ is the
angle of source-lens relative motion with respect to the direction of 
SIM-Earth axis at the moment when the event is a maximum as seen
from the SIM-Earth midpoint.  It is
only by measuring {\it both} of these quantities that one can determine
$\tilde r_\E$.  Since $\Delta\beta$ is difficult to measure, 
obtaining a precise $\tilde r_\E$ is also difficult.  

	As discussed in the introduction, it is possible in principle to
break the degeneracy in $\Delta \beta$ photometrically if the 
photometry is good enough.  
We will show in \S\ 5 that SIM photometry is sufficiently
precise for this task provided that the observations are carefully planned.
However, it is also the case that SIM astrometric measurements
{\it by themselves} often determine $\phi$ (from the orientation of the 
ellipse) with sufficient precision to break the degeneracy in $\Delta\beta$.
In these cases, $\tilde r_\E$ can be determined from a measurement of
$\Delta t_0$ alone.  In this paper, we will consider both methods of
breaking the degeneracy, but in the remainder of
this section we will assume that the degeneracy is broken astrometrically.
This will allow us to estimate $\Gamma_{\rm ph/ast}$, the relative precision
of SIM photometric parallax measurements to SIM astrometric
proper-motion measurements.

	As currently designed, SIM will fly in a SIRTF-like
orbit, drifting away from the Earth at about 0.1 AU per year.  Let the 
distance
at the time of the observations be $d_{\oplus-s}$.  Then $\tilde r_\E$ can
be determined from the measured $\Delta t_0$ (and the known value of $\phi$)
by,
\begin{equation}
\tilde r_\E = d_{\oplus-s}{t_\E\over |\Delta t_0|}|\cos\phi|.
\label{eqn:retildedet}
\end{equation}
Gould (1999a) analyzed how to optimize measurements of $\Delta t_0$ when he 
investigated microlens parallaxes with SIRTF.
For photon-limited photometry,
one should concentrate the measurements near times $t_\pm$ before and after the
peak, where $t_\pm = t_0 \pm (5/3)^{1/2}\beta t_\E$.  The error in $t_{0,s}$
is then approximatley given by
\begin{equation}
\sigma_{t_{0,s}} \sim n^{-1/2}\sigma_{\rm ph}\,{\beta\over 0.5}t_\E\qquad
(\beta\la 0.5),
\label{eqn:sigtnaught}
\end{equation}
where $n$ is the total number of measurements in these two regions.  We will
assume that the ground-based measurements to determine $t_{0,\oplus}$
have similar precision as the SIM measurements.  Equation
(\ref{eqn:retildedet}) then implies
that the fractional error in $\tilde r_\E$ is given by,
\begin{equation}
{\sigma_{\tilde r_\E}\over \tilde r_\E} = 
2^{1/2}{\sigma_{t_{0,s}}\over |\Delta t_0|}
=2^{1/2}{\tilde v\sigma_{t_{0,s}}\over d_{\oplus-s}}|\sec\phi|.
\label{eqn:sigretilde}  
\end{equation}
Then, assuming
that a similar number of measurements are used to determine the semi-major
axis of the ellipse, $\theta_a$, and to determine $\Delta t_0$ (they
are somewhat the same measurements), the ratio of the {\it photometric} 
precision
of the parallax to the {\it astrometric} precision of the proper motion is
\begin{equation}
\Gamma_{\rm ph/ast} = {\sigma_{\tilde r_\E}/\tilde r_\E\over
\sigma_{\theta_a}/\theta_a}
\sim {\tilde r_\E\sigma_{\rm ph} (\beta/0.3)|\sec\phi|/d_{\oplus-s}
\over \sigma_\theta/(\theta_\E/3)},
\label{eqn:gammaphastone}
\end{equation}
where we have approximated equation (\ref{eqn:thetaab}) 
as $\theta_a=\theta_\E/3$.
Using equations (\ref{eqn:msolve}) and (\ref{eqn:sigrel}), this can be 
rewritten
\begin{equation}
\Gamma_{\rm ph/ast} \sim \beta {4 G M\over c^2 d_{\oplus-s}\theta_f}
|\sec\phi| \sim {\beta\over 0.25}\,{M\over 0.3 M_\odot}
\biggl({d_{\oplus-s}\over 0.2\,\rm AU}\biggr)^{-1}|\sec\phi|,
\label{eqn:gammaphasttwo}
\end{equation}
where we have adopted $\theta_f=2.5\,$mas, which is appropriate if the 
flux-weighted harmonic mean wavelength of the source is $0.8\,\mu$m, and
the mirrors are separated by 10 m.
Equation (\ref{eqn:gammaphasttwo}) implies that for typical lenses, the
photometric parallax will be of comparable precision to the 
astrometric proper motion in sharp constrast to the large ratio
for astrometric parallaxes found in equation (\ref{eqn:relerrsast}).

	In fact, the $|\sec\phi|$ dependence in equation 
(\ref{eqn:gammaphastone}) is too pessimistic because we have ignored all
photometric information about $\Delta \beta$.  We show in \S\ 5 that except
for the case $\cos\phi\simeq 0$ (where the discrete degeneracy becomes
astrometrically
incorrigible) it is possible to essentially eliminate the $|\sec\phi|$ term in 
equation (\ref{eqn:gammaphastone}) using a combination of astrometric and
photometric data.

\section{Simulations}

	The estimates given in the previous two sections are useful because
they elucidate the relation between the physics of the event and the
measurement process on the one hand and the errors in the microlensing 
parameters on the other.  By the same token, however, they cannot capture
the full range of experimental conditions, and so are necessarily approximate.
The actual errors for any given event will depend both on the precise
event parameters and on the observational strategy.  While a full investigation
of the best observational strategy lies well beyond the scope of the present
study, it is important to make a rigorous calculation of the statistical
errors for some representative examples in order to obtain more precise
estimates and to investigate more subtle effects that are not captured by
the rough analysis given above.

	For this purpose, we consider a set of somewhat idealized 
observations.  First, we assume that the principal measurements are carried out
at uniform time intervals that are short compared to $t_\E$,
beginning when the magnification first reaches $A=1.5$ and ending at
a time that will be determined below from signal-to-noise (S/N) considerations.
This is quite reasonable for observations toward the LMC (near the ecliptic
pole) but is obviously not really possible toward the bulge (near the
ecliptic).  For the bulge, we therefore assume that the measurements are
interrupted when the bulge is within $60^\circ$ of the Sun.  This measurement
strategy can actually be very far from ideal, and we modify it somewhat 
in \S\ 4.2, below.  Second, we assume that the exposure times are of equal
durations, so that the S/N is better near the peak of the event.
Third, we assume $\theta_f=2.5\,$mas and $d_{\oplus-s}=0.2\,$AU, although
the first will clearly vary from star to star, and the second will change
during the course of the mission.  For the LMC (at the ecliptic pole), the
time of year at which the event is discovered plays no role, but for
the bulge it does.  We will assume that the bulge field lies at $B=-6^\circ$
from the ecliptic, 
close to the (northern) winter solstice.  The Earth-satellite
separation projected onto the sky is therefore at a maximum at the summer
solstice and is 
$d_{\oplus-s}|(1-\cos^2 \psi\cos^2 B)]^{1/2}\sim d_{\oplus-s}|\sin\psi|$ 
at other times of
the year, where $\psi$ is the phase of the Earth's orbit relative to the
autumnal equinox.  As discussed in detail by Gaudi \& Gould (1997), the 
orbital phase $\psi$ has two conflicting effects.  First, when $\cos\psi\sim 0$
(near the summer solstice), the SIM-Earth projected separation is a 
maximum, and hence the measurement errors of $\Delta t_0$ and $\Delta \beta$
are reduced to a minimum.  On the other hand, the relative velocity of 
SIM and the Earth projected onto the plane of the sky is at a minimum,
and breaking the degeneracy in $\Delta\beta$ depends critically on this
relative velocity.  Hence, it is most difficult to break the degeneracy
at the summer solstice.  As the phase moves away from the summer solstice,
the projected separation slowly declines (making the measurements of 
$\Delta t_0$ and $\Delta\beta$ less accurate) but the projected velocity
difference rapidly increases (allowing more secure degeneracy breaking).
If one relies on photometry to break the degeneracy, the optimal events are
those that peak about 45 days from the summer solstice (see e.g. Fig.\ 6 from
Gaudi \& Gould 1997).  On the other hand, if it is possible to break the
degeneracy astrometrically, then events that peak at the solstice are optimal
since the measurement errors are reduced by $\sim 2^{-1/2}$.  In this paper,
we will primarily simulate events peaking at $\psi = 225^\circ$, i.e. about
May 7.  However, we will also discuss events that peak at the summer solstice
($\psi = 270^\circ$).

Next, we will assume for definiteness that the ground-based photometric
observations have the same precision as the SIM observations.
	Finally, we will ignore blending in the SIM measurements,
except blending by the lens.
Blending will have an important impact on the overall precision and hence
the strategy for SIM measurements but as we show below, it will not
affect the main conclusions of this paper which concern the relative
precision of SIM astrometric and photometric microlens parallaxes.  
This is especially so toward the bulge which will be the main focus of 
analysis.  A proper treatment of blending would therefore make the paper 
substantially more complex without clarifying any of the central points.  
Hence, we defer consideration of this important effect to a later paper on 
observational strategy.

	Why can blending be ignored to first order in this analysis?  First,
Han \& Kim (1999) have shown that all potential blends lying
more than 10 mas from the source can be eliminated by the SIM
observations themselves.  Since the density of field stars having even
a modest fraction of the source flux is much less than $10^4\,\rm arcsec^{-2}$,
this essentially eliminates all blends not directly associated with the
event, namely the lens itself, binary companions to the lens, and binary
companions to the source.  For bulge events, 10 mas corresponds to 80 AU,
so a substantial fraction of binary companions are also eliminated.
Second, to minimize observation time, SIM observations must be
almost entirely restricted to very bright stars.  This means clump giants
toward the bulge and either clump giants or early main-sequence stars toward
the LMC.  The chance that a bulge clump star has a companion within 80 AU and 
with more than a few percent of its own flux is small because their 
progenitors are about 50 times fainter than the stars themselves.  The 
primary effect of a few percent blend would be to change the shape and
orientation of the proper-motion ellipse and (assuming the shape change
went undetected) 
to therefore change the inferred direction of the lens-source
relative proper motion, also by a few percent.  This would in turn affect the
parallax inferred from $\Delta t_0$ which depends on this direction through
the angle $\phi$.  See equation (\ref{eqn:retildedet}).  However, this effect
will also be a few percent.  As we will show in \S\ 5, it is quite possible to 
achieve accuracies of a few percent for bulge events, so a careful 
investigation of the effect of blending on parallax and proper-motion
measurements should be undertaken as part of a more thorough analysis
of the problem.  Unfortunately, a proper analysis of blending from binary
companions to the source requires simulated fits to the entire diffraction 
pattern, not just the centroid, and so is substantially more involved than
the present study.  By contrast, low-level blending by the lens can be
treated within the framework of the 
centroid analysis given here and we therefore include it.

	The situation is more complicated toward the LMC because the chance 
that an early main sequence star has a companion of comparable brightness is 
larger, probably a few tens of percent.  Even clump stars have brighter
companions toward the LMC than toward the bulge because they are younger and
so have brighter progenitors.  Also, the 10 mas limit on detecting blends
directly (Han \& Kim 1999) corresponds to 500 AU toward the LMC compared to
80 AU toward the bulge.  Nevertheless, even toward the LMC, the majority
of sources will not have companions with more than 10\% of the source flux
and hence even here it is appropriate to ignore blending by companions
in a first treatment.

	Toward both the bulge and the LMC, it is very unlikely that the
lens itself will contribute more than a small fraction of the source light
if the source is bright.  We therefore allow for lens blending in our simulated
fits but assume that the actual blending is very small.

	Note that we will {\it not} ignore blending in the ground-based
photometric observations, since there is no way to eliminate field
star blends for the ground-based observations.  

\subsection{Parameterization}

	We will simulate simultaneous observations from SIM and
from the ground.  There will be four measured quantities: 1) $G^1$, 
the flux observed from the ground, 2) $G^2$, the flux observed from SIM, 
3) $G^3$, the $x$ astrometric position, 
and 4) $G^4$, the $y$ astrometric position.  These give rise to four
observational equations,
\begin{equation}
G^1(t) = F^1_s A(u_\oplus) + F^1_b, \qquad
G^2(t) = F^2_s A(u_{SIM})  + F^2_b
\label{eqn:goneandtwo}
\end{equation}
and
\begin{equation}
[G^3(t),G^4(t)] =   {{\bf u}_{SIM}\over u^2_{SIM} + 2}\theta_{\rm E}
- \balpha_{SIM}\pi_s + \bmu_s t + \btheta_0 - \btheta_b
\label{eqn:gthreeandfour}
\end{equation}
\begin{equation}
\btheta_b \equiv  {F^2_b\theta_\E\over
F_s^2 A(u_{SIM}) + F_b^2}
\biggl[{u^2_{SIM} + 3\over u^2_{SIM} + 2}{\bf u}_{SIM} + \kappa\balpha_{SIM}
\biggr]
\label{eqn:thetasb}
\end{equation}
where $A(u)=(u^2+2)/[u(u^2+4)^{1/2}]$ is the magnification,
$\bmu_s$ is the absolute proper motion of the source, $\btheta_0$ is
the true position of the source at time $t=0$,
\begin{equation}
\balpha = -{{\bf \hat n}\times {\bf \hat n}\times {\bf a}\over {\rm AU}},
\label{eqn:balphadef}
\end{equation}
$\pi_s$ is the parallax of the source, 
and ${\bf u}_{SIM}$ and ${\bf u}_\oplus$
are defined similarly to equation (\ref{eqn:pardev}), i.e.
\begin{equation}
{\bf u}_{SIM} = {\bf u}_\odot + \kappa\balpha_{SIM},\qquad 
{\bf u}_{\oplus} = {\bf u}_\odot + \kappa\balpha_{\oplus},\qquad 
\kappa\equiv
{{\rm AU}\over \tilde r_\E},
\label{eqn:usimdef}
\end{equation}
with
\begin{equation}
{\bf u}_\odot \equiv (\tau_\odot\cos\phi-\beta_\odot\sin\phi,
\tau_\odot\sin\phi + \beta_\odot\cos\phi),
\qquad \tau_\odot\equiv {t-t_{0,\odot}\over t_{\E,\odot}}.
\label{eqn:udotdef}
\end{equation}
The terms in equation (\ref{eqn:gthreeandfour}) can be understood as follows.
The first term is the ellipse characterized by equation (\ref{eqn:thetaab}),
modified by the motion of the Earth (i.e. ${\bf u}_\odot
\rightarrow {\bf u}_{SIM}$).
The second term is the parallactic motion of the source.  The third and
fourth terms represent the ordinary proper motion and position of the source.
The last term is the perturbation due to the luminosity of the lens which
is written out explicitly in (\ref{eqn:thetasb}).  Here the first term
is the difference between the ``ellipse'' in equation 
(\ref{eqn:gthreeandfour}) and the relative lens-source position 
($-{\bf u}_{SIM}\theta_\E$), 
while the second is the relative parallax of the lens
and the source.

	There are a total of 15 parameters: $t_{0,\odot}$, 
$\beta_\odot,$ and 
$t_{\E,\odot}$ 
are the standard event parameters as it would be seen from the Sun,
$\phi$ is the direction of source motion relative to the lens with respect
to the SIM-Earth direction,
$\theta_\E$ is the Einstein radius, $\kappa$ is the inverse projected
Einstein radius (normalized in AU), $\pi_s$ is the parallax of the source,
$\bmu_s$ is the proper motion of the source, $\btheta_0$ is its position at
$t=0$, $F^1_s$ and $F^2_s$ are the source fluxes as received by the Earth
and satellite observatories, and $F^1_b$ and $F^2_b$ are the background 
fluxes.

	To determine the uncertainties in these parameters, we evaluate
the covariance matrix $c_{i j}$ (e.g. Gould \& Welch 1996)
\begin{equation}
c = b^{-1},\qquad b_{i j} = \sum_{l=1}^4\sum_{k=1}^N
\sigma_{k l}^{-2}{\partial G^l(t_k)\over \partial a_i}
{\partial G^l(t_k)\over \partial a_j},
\label{eqn:cijdef}
\end{equation}
where $a_1$ ... $a_{15}$ are the fifteen parameters, $t_k$ are the times
of the observations, and $\sigma_{k l}$ is the error in the measurement of
$G^l$ at time $t_k$.  We enforce the condition of weak blending by setting
$F^1_b=F^2_b=0$ {\it after} taking the derivatives in equation 
(\ref{eqn:cijdef}).

\subsection{Modification of observing strategy}

	The full problem of optimization of SIM microlensing observations
lies outside the scope of this paper, but it is straight forward to determine
the optimum duration of observations once the (arbitrary) strategy of
uniform observations has been adopted:  one changes the interval over which 
the observations are carried while holding the total observing time fixed
and inspects the resulting errors.
We carry out this exercise and find that the optimal duration to determine
$\tilde r_\E$ is short, typically a few tens of days for various combinations
of parameters, while the optimal duration to determine $\theta_\E$ is well
over 100 days.  The reason for this is clear.   The measurement of
$\tilde r_\E$ is determined primarily from photometry, and the photometric 
microlensing event is essentially over after $2 t_\E$.  On the other hand 
$\theta_\E$ is determined from the astrometric event which lasts many
$t_\E$.  Only after the astrometric event is essentially over is it possible
to determine $\bmu_s$ and so remove the correlation between this parameter
and $\theta_\E$.  

	We address this inconsistency of time scales by
 modifying the observational strategy.
We take observations uniformly over various intervals but with fixed
total observing
time, $T$, and then take three additional observations, each with observing
time $T/20$, at 3 months, 9 months, and 12 months after the peak of the 
event.  Thus, the total observing time is $1.15\,T$.  We then find that
the precisions of both $\tilde r_\E$ and $\theta_\E$ are roughly constant when
the continuous observations last anywhere from 30 to 120 days.  For
definiteness, we adopt 50 days uniformly.

\subsection{S/N: assumptions and scalings}

	We assume that all sources are $I=15$ which is typical of the
brighter microlensing events seen toward the bulge but is much brighter
than any event seen so far toward the LMC.  In our calculations, we 
normalize the astrometric precision by assuming that $4\,\mu$as accuracy
(in one dimension) can be achieved in 1 minute of observation
on an $I=11$ star.  That is, our fiducial $I=15$ stars require 40 minutes
to reach $4\,\mu$as.  We allow a total of 5 hours of observation for each
event.  It is then straight forward to scale our results to other assumed
conditions.  For example, for an $I=20$ source, the errors reported in the
next section must be multiplied by 10.  Alternatively, the same errors could
be achieved by allowing 500 hours of observations.  If our astrometric
error estimate proves too pessimistic, so that it is possible to achieve
$4\,\mu$as precision in a minute on an $I=12$ star, then the errors should
be divided by 1.6.

\section{Results}

\subsection{Bulge}

	We consider two geometries.  The first is a bulge line of sight
at $6^\circ$ from the ecliptic.  The source and lens are both in the bulge,
with $D_{\rm os}=8\,$kpc and $D_{\rm ol}=6\,$kpc.  Hence $D=24\,$kpc.
(See eq.\ \ref{eqn:thetae}.)
The speed of the lens relative to the observer-source line of sight is
$v=200\,\kms$.  We vary $M$, $\phi$, and $\beta$.  Formally $\beta$ is
defined as the impact parameter of the event as seen from an 
observer at the Earth-SIM midpoint but in practice it
is very similar to the $\beta$ observed from the Earth.  All events are assumed
to peak (as seen from the midpoint) on May 7, i.e., 45 days before
the (northern) summer solstice.  This is the most favorable time to 
break the discrete degeneracy in $\Delta\beta$ photometrically
(see Gaudi \& Gould 1997), but the intrinsic errors in $\Delta\beta$
and $\Delta t_0$ are larger by $\sim 2^{1/2}$ than they would be at
the summer solstice.  We will therefore later investigate whether the
discrete degeneracy in $\Delta\beta$ can be broken astrometrically so that
observations could take place at the solstice.  In order
to better understand this and several other issues, we conduct two sets of
simulations, one using SIM measurements alone (both astrometric and 
photometric) and the other combining astrometry
with photometry from both SIM and the ground.

	Table 1 shows the results.  The first three columns are the input 
parameters $M$, $\beta$, and $\phi$.  Columns 4 and 5 show the errors in
$\phi$ from SIM measurements only and with the addition of 
ground-based measurements. Columns 6 through 11 show the fractional errors
in various quantities both without and with ground-based photometry.
Columns 6 and 7 show the errors in
$\tilde r_\E$,
columns 8 and 9 show the errors
in $\theta_\E$, and
columns 10 and 11 the errors 
in $M=\tilde r_\E\theta_\E c^2/4 G$.  Finally, columns 12 and 13 show the
fractional errors
in $D=\tilde r_\E/\theta_\E$, and in $\pi_s$ based on combined data from
SIM and the ground.

	These results are in rough qualtitative agreement with the predictions
of equations (\ref{eqn:relerrsast}) and (\ref{eqn:gammaphasttwo}).  In
particular, they confirm that the fractional error in $\tilde r_\E$ is
orders of magnitude larger than the fractional error in 
$\tilde \theta_\E$ if one is restricted to SIM data, while the
two errors are comparable if one combines astrometric with photometric data.

	However, there are a number of additional important conclusions that
can be drawn from Table 1.  First, restricting consideration to impact
parameters $\beta<0.5$, the fractional errors in $M$, $D$, and $\pi_s$
are all typically about 5\% although these errors do vary
somewhat in particular cases.  This means 5 hours of observation produce
very precise individual solutions for the mass, distance, and velocity of the
lens, and also for the distance and velocity of the source,
implying that a few hundred
hours of SIM time could yield a very detailed inventory of the material
between the Sun and the Galactic center.

	Second, while the errors in $\tilde r_\E$ do deteriorate 
toward $\phi=0$, the trend is not as drastic as predicted by
equation (\ref{eqn:gammaphasttwo}).  The fundamental reason for
this is that the continuous degeneracy in $\Delta \beta$ is not very severe,
so that if one assumes that the discrete degeneracy is broken, then there
is actually quite a lot of information about this component of $\tilde r_\E$
in the photometric measurements.  In fact, the values in Table 1 implicitly
{\it do assume} 
that the discrete degeneracy is broken.  This is because they are
based on equation (\ref{eqn:cijdef}) which gives a {\it purely local}
error analysis.  

	Recall that there are two discrete degeneracies.  We focus initially 
on the one that affects the {\it magnitude} of 
$\Delta\beta$ and defer consideration of the one that affects only 
the sign of $\Delta\beta$.
To determine whether this discrete degeneracy is broken photometrically,
we examine the work of Gaudi \& Gould (1997), in particular their Figure 6.
Under the observational conditions they considered, the discrete degeneracy
is broken 90\% of the time for $M=0.3\,M_\odot$, 
$\psi=225^\circ$, and satellite separations
$d_{\oplus-s}=0.4\,$AU.  This is twice the separation that we have assumed.
However,  Gaudi \& Gould (1997) 
have assumed photometry errors of 1\% for the Earth and
2\% for the satellite for a total of about 70 observations.  If our 5 hours
of observing time were divided among 70 observations, the photometric 
precision would be 1.2\%.  That is, our assumed effective errors are smaller
by a factor $[(1.2^2 + 1.2^2)/(1^2 + 2^2)]^{1/2} = 0.75$.  For small
Earth-satellite separations, there is a direct trade off between
measurement error and satellite separation, so our 0.2 AU separation 
corresponds to 0.25 AU in their simulations.  
Inspection of the Gaudi \& Gould (1997) Figure 6 shows
that the degeneracies
would be broken 70\% of the time.  To determine how this effectiveness
scales with lens mass, we turn to Figure 4 from Gaudi \& Gould (1997). 
This shows that degeneracy breaking becomes more difficult at lower masses.
The figure is drawn for the case $d_{\oplus-s}=1\,$AU whereas the argument
just given implies that with 5 hours of observation, our 0.2 AU separation
is equivalent to 0.25 AU in the Gaudi \& Gould (1997) simulations.  Hence,
comparing Gaudi \& Gould (1997) Figures 4 and 6, we estimate that 35\%
of the degeneracies would be broken for $M=0.1\,M_\odot$.  If the exposure
times were multiplied 
by a factor of 4 to 20 hours, this fraction would rise to about
70\%.

	We now turn to the question of how well the degeneracies can be
broken astrometrically.
	For the geometry considered here, 
$\tilde r_\E= 7.5(M/0.3\,M_\odot)\,$AU.  Hence $\Delta \beta \sim
d_{\oplus-s}/\tilde r_\E\sim 0.025$, which is quite small compared to typical
values of $\beta$.  This means that for most events 
the discrete degeneracy will
be between solutions with $\Delta\beta\sim 0.025$ and $\Delta\beta\simeq 2\beta
\sim 0.5$.  Since $\tan\phi=\Delta \beta/\Delta t_0$, the high $\Delta\beta$
solution will almost always correspond to angles $90^\circ\pm 2^\circ$
or $270^\circ\pm 2^\circ$,
while the low $\Delta\beta$ solution (assuming that it
is the real one) will be at some random angle.  Thus to distinguish the
two solutions, one must have independent information about $\phi$ with
errors that are a factor $\sim 3$ smaller than $|90^\circ-\phi|$ or
$|270^\circ-\phi|$.  Is $\phi$
this well constrained by the observations?  Looking at column 4 of Table 1,
$\phi$ seems to be very well constrained.  However, this precision measurement
is based primarily on the photometric measurements of $\Delta\beta$ and 
$\Delta t_0$ and so implicitly assumes that the discrete degeneracy has
been broken.  Hence we should use only SIM data (column 5).

	We see from column 5 of Table 1 that the errors in $\phi$ are 
generally small for $\phi\leq 60^\circ$ but
deteriorate toward $\phi=90^\circ$.  This means that the discrete
degeneracy is broken astrometrically at the $3\,\sigma$ level 
for $\phi\leq 60^\circ$ but cannot
be broken if the angle gets close enough to $90^\circ$.  Table 1
does not have sufficient resolution to determine the transition but we find
by more detailed calculations that for $\beta = (0.2,0.4)$ this occurs at
($70^\circ$,$65^\circ$), for $M=0.1\,M_\odot$, at
($75^\circ$,$70^\circ$), for $M=0.3\,M_\odot$, and at
($80^\circ$,$75^\circ$), for $M=0.5\,M_\odot$.  Hence, the degeneracy is
usually broken astrometrically, but less frequently at low masses.
Since the degeneracy is more difficult to break at low masses both
photometrically and astrometrically, it would be prudent to commit more
observation time (say 20 hours rather than 4 hours) to the shortest
events (which are likely to be low mass).  

	The fact that the degeneracy can be broken astrometrically for
most events seems to argue against restricting observations to periods
that are 45 days from the summer solstice.  Recall that we adopted this
restriction in order to permit better photometric degeneracy breaking which
now no longer seems so necessary.  However, we find that for events peaking
at the solstice, the errors in $\phi$ (when the Earth-based observations
are ignored) are substantially higher
than the values in column 5, implying that
it is not often not possible to break the discrete degeneracy astrometrically
or photometrically at the solstice.

	As we have noted, the above discussion actually applies to only one
of two discrete degeneracies, the one involving two different {\it magnitudes}
of $\Delta \beta$.  This is the more important degeneracy because it affects
the estimate of the size of $\tilde r_\E$ and so of the mass, distance, and
speed of the lens.  However, there is also another degeneracy involving the
{\it sign} of $\Delta\beta$ but not its magnitude.  Boutreux \& Gould (1996)
and Gaudi \& Gould (1997)
refer to the first of these as the ``speed degeneracy''.  We call the second
the ``direction degeneracy''.  The direction
degeneracy becomes difficult to break when 
$|\Delta\beta|\ll |\Delta t_0|/t_\E$, i.e., when $\phi\sim 0^\circ$
or $\phi\sim 180^\circ$, so that the two degenerate solutions are close
in $\phi$.  From column 5 in Table 1, we find that the error in $\phi$ from
the astrometric data alone is generally quite small for $\phi=0$,
and hence is adequate to break the direction degeneracy unless $\phi$ lies
within a few degrees of either $0^\circ$ or $180^\circ$.  However, in this
case the effect of the degeneracy is very small.

	Finally, we note that we have examined the errors in $F_b/F_s$
(although we do not display them).  We find that they are typically a few
percent, implying that a luminous lens could be detected if it were more
than a few percent of the flux of the source.

\subsection{LMC}

The second line of sight is toward the LMC at the south ecliptic pole.
The source lies at $D_{\rm os}=50\,$kpc, while the lens is assumed to be
in the halo at $D_{\rm ol}=15\,$kpc.  Hence $D=21\,$kpc, which is very
similar to the bulge value.  This means that at fixed mass, the bulge
events considered in \S\ 5.1 will have about the same $\tilde r_\E$ as the
LMC events considered here.
The speed of the lens relative to the observer-source line of sight is
$v=250\,\kms$, slightly larger than for the bulge.
We have again assumed an $I=15$ source to allow easy 
comparison with results from the bulge.  We discuss the implications of this
highly unrealistic assumption below.

	Table 2 shows the results.  They are qualitatively similar to those
for the bulge.  The largest difference is that the fractional error in $\pi_s$
is larger which simply reflects the fact that the LMC is more distant.

	Of the eight microlensing events detected by Alcock et al.\ (1997b)
during their first two years of observations, none of the sources were 
brighter than $V=20$ (after removing blended light) which is the nominal
limit for SIM.  Future microlensing surveys could improve the rate of
detection by an order of magnitude (Gould 1999b; Stubbs 1998).  However,
only a factor of three of this improvement would be 
due to the coverage of a larger
area.  The rest would come from going deeper which would not yield any more
bright sources.  Hence, the total rate of events that are accessible to
SIM will not be high.  Most of the usuable events that are detected
are likely to be close to the magnitude limit.  For astrometric
measurements $V=20$ is equivalent to $I=19$ because at fixed magnitude there
are fewer photons in $I$ and they are longer.  Hence, for the same total
observing time (5 hours), the errors will be about a factor of $40^{1/2}$
larger than listed in Table 2.  That is, the mass and distance estimates will 
be accurate to about 20 to 40\%.  This would be an acceptable level of
precision to resolve the
question of the nature of the lenses assuming that more than a handful of
events can
be measured.  The errors in the measurement of $\phi$ without making use of
the Earth-based observations (column 5) are typically $15^\circ$
(after including the factor $40^{1/2}$).  Hence in many cases
it will not be possible
to break the $\Delta\beta$ degeneracy astrometrically.

	To determine whether the photometry is sufficiently precise 
to break the
degeneracy, we compare our simulation with that of Boutreux \& Gould
(1996) who specifically considered an Earth-satellite separation of 0.26 AU
which is close to our value of $d_{\oplus-s}=0.2$ AU.  
In their Monte Carlo simulation, the 
``speed degeneracy'' (between different scalar values of $\Delta \beta$)
was broken 40--60\% of the time in  the mass range 0.1--1 $M_\odot$.
We find that their assumed photometric errors are about equal to
the ones assumed here (for a total of 5 hours of observation at $V=20$).
Therefore, it seems likely that the SIM photometric observations
would be adequate to break this degeneracy in the majority of cases.

	Inspection of column 5 from Table 2 shows that the direction
degeneracy will usually be broken astrometrically.
The simulations of Boutreux \& Gould (1996)
show that it is about equally difficult to break the direction
degeneracy as the speed degeneracy.  Thus, it should usually also 
be possible to break this degeneracy photometrically.  In any event,
as in the case of the bulge, the direction degeneracy is much less 
important than the speed degeneracy.

	While 5 hours of observation is adequate to characterize events
toward the LMC, it would seem advisable to invest more time for these
events.  First, there will be very few of them so the cost in total observing
time is not high.  Second, if the events are in the halo, one would like
to know their mass more accurately than 30\% if possible.  Finally, if the
lenses reside in the LMC, then 5 hour exposures would yield only a 
non-detection of parallax rather than a measurement.  While this 
(together with the proper-motion measurement) would be highly significant
in demonstrating that the lenses were not in the halo, it would provide
only limited clues as to their nature.  In brief, the scientific return
of extra observation time would be high for LMC events.

{\bf Acknowledgements}: We thank Scott Gaudi for his careful comments on
the manuscript.  After this manuscript was almost complete, we learned of
closely related work by Cheongho Han.
This research was supported in part by grant AST 97-27520 from the NSF.

\end{document}